\documentclass[letter,scriptaddress,twocolumn, pre,showkeys,showpacs]{revtex4-1}

	\usepackage{amsmath}
	\usepackage{makeidx}
	\usepackage{amsfonts}
\usepackage{amssymb}
\usepackage{MnSymbol}
	\usepackage[ansinew]{inputenc}
	\usepackage[usenames,dvipsnames]{pstricks}
	\usepackage{subfigure}
	\usepackage{epsfig}
	\usepackage{pst-grad} 
	\usepackage{pst-plot} 
	\usepackage[colorlinks,hyperindex]{hyperref}
	\hypersetup
	{
		colorlinks,%
		citecolor=black,%
		linkcolor=black,%
		urlcolor=black,%
	}

\makeindex

\usepackage{tabulary}
\usepackage{color,soul}
\definecolor{myyellow}{rgb}{1,1,0}
\definecolor{myred}{rgb}{1,0,0}
\definecolor{myblue}{rgb}{0,0,1}
\definecolor{gray}{rgb}{0.5,0.5,0.5}
\sethlcolor{myyellow}
\setstcolor{myred}
\setulcolor{myblue}

\begin{document}
\title{The effects of fibroblasts on wave dynamics in a
mathematical model for human ventricular tissue}

\author{Alok Ranjan Nayak$^{1}$}
\email{alok@cps.iisc.ernet.in}
\author{Rahul Pandit$^{2}$}
\email{rahul@physics.iisc.ernet.in}
\affiliation{
$^1$Robert Bosch Centre for Cyber Physical Systems,
Indian Institute of Science, 
Bangalore 560012, India.\\
$^2$Centre for Condensed Matter Theory,
Department of Physics, Indian Institute of Science,
Bangalore 560012, India.}



\begin{abstract}{
We present systematic numerical studies of electrical-wave
propagation in two-dimensional (2D) and three-dimensional (3D)
mathematical models, for human, ventricular tissue with myocyte
cells that are attached (a) regularly and (b) randomly to
distributed fibroblasts.  In both these cases we show that there
is a parameter regime in which single rotating spiral- and
scroll-wave states (RS) retain their integrity and do not evolve
to a state ST that displays spatiotemporal chaos and turbulence.
However, in another range of parameters, we observe a transition
from ST to RS states in both 2D or 3D domains and for both cases
(a) and (b).  Our studies show that the ST-RS transition and
rotation period of a spiral or scroll wave in the RS state
depends on (i) the coupling strength between myocytes and
fibroblasts and (ii) the number of fibroblasts attached to
myocytes. We conclude that myocyte-fibroblast coupling strength
and the number of fibroblasts are more important for the ST-RS
transition than the precise way in which fibroblasts are
distributed over myocyte tissue.}
\end{abstract}



\maketitle

\section{Introduction}
\label{intro}

Approximately $16\%$ of all deaths in the industrialized world
are caused by cardiac arrhythmias like ventricular tachycardia
(VT) and ventricular fibrillation (VF)~\cite{vf:1,vf:2}. There is
a broad consensus that the analogs of VT and VF in mathematical
models for cardiac tissue are, respectively, (a) a single
rotating spiral or scroll wave of electrical activation and (b)
spiral-wave or scroll-wave turbulence, which displays broken
electrical waves and spatiotemporal
chaos~\cite{davidenko:1992,pertsov:1993,gray:1998,jalife:1998}.
Thus, it is very important of study such spiral and scroll waves
to develop an understanding of life-threatening cardiac
arrhythmias. Such studies are truly interdisciplinary because
they require inputs from biology, bio-medical engineering,
cardiology, on the one hand, and physics, nonlinear dynamics, and
numerical methods, on the other. We use  methods from these areas
to solve the nonlinear, partial-differential equations for
state-of-the-art mathematical models for cardiac tissue with
myocytes and fibroblasts.  We build on our earlier studies of
this problem~\cite{nayak:2013,nayak:pre} to elucidate the role of
two different forms of myocyte-fibroblast couplings on spiral-
and scroll-wave dynamics in such models by using theoretical
ideas from spatiotemporal chaos.

It is useful to begin with an overview of experimental and
computational studies of cardiac myocytes and fibroblasts.
Cardiac fibroblasts, which are inexcitable cells, often multiply
and connect with cardiac myocytes during
\textit{fibrosis}~\cite{krenning:2010,biernacka:2011,souders:2009},
a process of cardiac-tissue healing after a myocardial
infarction. Fibroblasts in cell culture and in intact tissue can
couple with myocytes by expressing either the connexins-43 (Cx43)
or 
Cx45~\cite{rook:1992,camelliti:2004,chilton:2007,zlochiver:2008,miragoli:2006}.
Zlochiver, \textit{et al.}~\cite{zlochiver:2008} have shown the
expression of Cx43  between fibroblasts and myocytes in a
monolayer of myocytes and fibroblasts of neonatal rats; Miragoli,
\textit{et al.}~\cite{miragoli:2006} have reported that Cx43 and
Cx45 are expressed among fibroblasts and between fibroblasts and
myocytes in cultured fibroblasts coated over
rat-ventricular-myocyte strands. Both experimental and
computational studies have shown that such coupling between
myocytes and fibroblasts enhances electrical-signal propagation
in cardiac
tissue~\cite{zlochiver:2008,mcspadden:2009,miragoli:2006,xie:2009,nayak:2013};
this enhancement increases with $N_f$, the number of fibroblasts
that are attached to a myocyte. In cell-culture experiments,
Miragoli, \textit{et al.}~\cite{miragoli:2006} have found that
the conduction velocity (CV) decreases with an increase in the
density of fibroblasts in cultured strands of neonatal-rat
myocytes coated by fibroblasts; studies by McSpadden, \textit{et
al.}~\cite{mcspadden:2009} have found that CV decreases as the
fibroblast number increases on the top of a myocyte layer in a
monolayer of neonatal-rat cardiac myocytes, which are
electrotonically loaded with a layer of cardiac fibroblasts.
Zlochiver, \textit{et al.}~\cite{zlochiver:2008} have shown that
CV decreases as (i) the gap-junctional conductance increases or
(ii) the fibroblasts density increases in their experiments with
fibroblasts of neonatal rats; they have also obtained similar
result in their computational studies in a two-dimensional (2D)
sheet of myocyte tissue in the dynamic Luo-Rudy (LRd)
model~\cite{lr2a,lr2b} by 
inserting fibroblasts.  Computational studies by Xie, \textit{et
al.},~\cite{xie:2009} have shown that CV decreases as they
increase the gap-junctional coupling or the fibroblasts density
in a 2D LR1 myocyte model~\cite{lr1}, with either randomly attached or
randomly inserted fibroblasts. In simulations with the 2D TNNP04
model (due to ten Tusscher, \textit{et al.}~\cite{tnnp04}), Nayak,
\textit{et al.}~\cite{nayak:2013} have found that CV either
decreases or increases, with attached fibroblasts, as they
increase the gap-junctional coupling. The experimental and
computational investigations mentioned above show that both the
gap-junctional coupling and $N_f$ enhance CV and, therefore, they
can play a crucial role in spiral- and scroll-wave dynamics in
mathematical models for cardiac tissue.

We develop and investigate two models with different arrangements
of fibroblasts that are attached to myocytes. In the first
arrangement there is a regular, spatially periodic attachment of
fibroblast, whereas, in the second arrangement, fibroblasts are
attached randomly to myocytes.  Our study has been designed to
understand the effects of fibroblast organization, fibroblast
density, and the myocyte-fibroblast coupling on spiral- and
scroll-wave dynamics.  We use two parameter sets for myocytes.
The first set leads to a stable rotating spiral or scroll (RS)
wave; the second leads to spiral- or scroll-wave-turbulence (ST)
states in an isolated myocyte domain. By investigating an ST state
in the presence of fibroblasts, we observe that both models, with
regularly and randomly attached fibroblasts, show transitions
from an ST to an RS state, depending on the myocyte-fibroblast
coupling $G_j$ and the maximum number $N_f$ of fibroblasts
attached to a myocyte in our simulation domain.  We find that,
once ST is converted to RS, the spiral or scroll rotation period
increases as we increase $G_j$ and $N_f$. Our study with an RS
state and fibroblasts shows that an RS remains unchanged in both
models with regularly and randomly attached fibroblast; and the
rotation period increases as we increase $G_j$ and $N_f$.

The remainder of this paper is organized as follows.
Section~\ref{modelmethod} is devoted to a description of our
model and the numerical methods we use. Section~\ref{result} is
devoted to our results.  Section~\ref{conclusion} contains a
discussion of the significance of our results.

\section{Model and Methods}
\label{modelmethod}

In this Section, we describe the details of our
myocyte-fibroblast models for two-dimensional (2D) and
three-dimensional (3D) tissue. We also explain the
numerical-simulation techniques that we use to solve the partial
differential equations (PDEs) that comprise our mathematical
models. We also discuss the methods that we use to analyze the
data from our numerical simulations.

\subsection{Model}
\label{model}

The 2D and 3D myocyte domains, with attached fibroblasts, can be
modeled by the following PDEs and ordinary-differential-equations
(ODEs)~\cite{keener:1998,panfilov:1997}:
\begin{eqnarray}
\label{eq:pde} 	
\frac{\partial V_m}{\partial t} &=& \frac{-I_m + N_f(\mathbf{x})I_j}{C_m} + D
\nabla^2 V_m,\\
\label{eq:ode} 	
\frac{\partial V_f}{\partial t} &=& \frac{-I_f - I_j}{C_f},
\end{eqnarray}
where
\begin{eqnarray}
	I_j=G_j(V_f-V_m);
\end{eqnarray}
here $C_m$ is the total cellular capacitance of a myocyte, $V_m$
is the myocyte transmembrane potential, i.e., the voltage
difference between intra- and extra-cellular spaces, and $I_m$ is
the sum of all the ionic currents that cross the myocyte cell
membrane; $C_f$, $V_f$, and $I_f$ are, respectively, the total
cellular capacitance, the transmembrane potential, and the sum of
all ionic currents for the fibroblast; $N_f(\mathbf{x})$ is the
number of identical fibroblasts attached to a myocyte in our
simulation domain at the point $\mathbf{x}$; and $I_j$, $G_j$,
and $D$ are, respectively, the gap-junctional current, the
myocyte-fibroblasts gap-junctional conductance, and the diffusion
coefficient that is related to the gap-junctional conductance
between myocytes.

For myocytes, we use the state-of-the-art mathematical model for
human ventricular tissue developed by ten Tusscher and Panfilov
(the TP06 model)~\cite{tp06}.  In the TP06 model the total ionic
current is
\begin{eqnarray}
I_m &=& I_{Na} + I_{CaL} + I_{to} + I_{Ks} + I_{Kr} + I_{K1}   \\ \nonumber
         & & + I_{NaCa} + I_{NaK} + I_{pCa} + I_{pK} + I_{bNa} + I_{bCa},
\end{eqnarray}
where $I_{Na}$ is the fast, inward ${Na^+}$ current, $I_{CaL}$
the L-type, slow, inward $Ca^{2+}$ current, $I_{to}$ the
transient, outward current, $I_{Ks}$ the slow, delayed, rectifier
current, $I_{Kr}$ the rapid, delayed, rectifier current, $I_{K1}$
the inward, rectifier ${K^+}$ current, $I_{NaCa}$ the
${Na^+}/{Ca^{2+}}$ exchanger current, $I_{NaK}$ the
${Na^+}/{K^+}$ pump current, $I_{pCa}$ and $I_{pK}$ the plateau
$Ca^{2+}$ and $K^+$ currents, and $I_{bNa}$ and $I_{bCa}$ the
background $Na^+$ and $Ca^{2+}$ currents, respectively. The full
sets of equations for this model, including the ODEs for the
ion-channel gating variables and the ion dynamics, are given in
Refs.~\cite{nayak:2013,thesis:alok}.

We follow MacCannell, \textit{et al.}~\cite{maccannell:2007} to
model the fibroblasts as passive elements. The fibroblast ionic
current $I_f$ is
\begin{equation}
\label{scell_fib}
I_f = G_f (V_f - E_f),
\end{equation}
where $G_f$ and $E_f$ are, respectively, the conductance and the
resting membrane potential for the fibroblast.

Physical units in our model are as follows: time $t$ is in
milliseconds (ms), $V_m$ and $V_f$ are in millivolts (mV), $C_m$
and $C_f$ are in picofarads (pF), $I_m$ and $I_f$ are in
picoamperes (pA), $G_f$ is in nanoSiemens (nS), $E_f$ is in mV,
$G_j$ is in nS, and $D$ is in cm$^2$/ms.

We study models with (a) regularly
attached fibroblasts and (b) randomly attached fibroblasts. In
case (a), $N_f(\mathbf{x})=N_f$ for all site $\mathbf{x}$ in our
simulation domain. In case (b) we choose $N_f(\mathbf{x})$
randomly at each site $\mathbf{x}$; $N_f(\mathbf{x})$ can be any
integer from 0 to $N_f$, with equal probability for any one of
these values.

\subsection{Methods}
\label{method}

Our 2D and 3D simulation domains are, respectively, squares
($1024\times1024$ grid points) and rectangular parallelepipeds
($1024\times1024\times8$ grid points). We use 5-point and
7-point stencils for the Laplacian in 2D and 3D, respectively,
and a finite-difference scheme with step sizes $\delta x = \delta
y = 0.25$~mm in 2D, and $\delta x = \delta y = \delta z =
0.25$~mm in 3D, i.e., our simulation domains are
$256\times256$~mm$^3$ (in 2D) and $256\times256\times2$~mm$^2$
(in 3D). For time marching we use a forward-Euler scheme with
$\delta t =0.02$~ms. We use Neumann (no-flux) conditions at the
boundaries of our simulation domain.

For numerical efficiency, we have carried out our simulations on
parallel computers, with an MPI code that we have developed for
the TP06 model. Our code divides the 2D (or 3D) simulation domain
into $n$ columns (or slabs) along the $x$-direction of the
domain, i.e., each processor carries out the computations for
$(1024/n) \times 1024$  and $(1024/n) \times 1024 \times 16$ grid
points, respectively, for 2D and 3D domains. To compute the
Laplacian at the interface of processor boundaries, we use two
extra grid lines (or surfaces), which can send and
receive the data from left- and right-neighbor processors. The
Neumann boundary condition is taken care of by adding an extra
layer of grid points on the boundaries of the simulation domain
of each processor. 

Reference~\cite{clayton:2008} suggests that we must have $D
\delta t/({\delta x}^2) < 1/2d$ for numerical stability, where
$d$ is the dimension of the simulation domain. For the TP06
model, with diffusion coefficients $D=
0.00154$~cm$^2$/ms~\cite{tp06}, time step $\delta t =0.02$~ms,
and space step $\delta x =0.25$~mm, the value of $D\delta
t/(\delta x)^2$ is $\simeq 0.05$; for the TP06 model, the
quantity $1/2d = 0.25$ and $\simeq 0.17$, for the 2D and 3D
domains, respectively, i.e., we have numerical stability because
$D\delta t/(\delta x)^2 < 1/2d$.  

We check the accuracy of our numerical scheme, as in
Ref.~\cite{clayton:2008}, by varying both  $\delta t$ and $\delta
x$ in a cable-type domain of
myocytes~\cite{thesis:alok,nayak:2013} and by measuring CV of a
plane wave, which is injected into the domain by stimulating its
left boundary for $3$~ms with a stimulus of strength $150$~pA.
With $\delta x=0.25$~mm, CV increases by $1.1\%$ when we change
$\delta t$ from $0.02$ to $0.01$ ms; if we decrease $\delta x$
from $0.25$ to $0.15$~mm, with $\delta t=0.02$~ms, then CV
increases by $3.1\%$; these changes are similar to those found in
other
studies~\cite{tnnp04,clayton:2008,shajahan:2009,thesis:alok}.

Although the numerical method we use satisfies both
numerical-stability and accuracy conditions, an inappropriately
large  $\delta x$ can give irregular wavefront-curvature, as a
consequence of numerical
artifacts~\cite{clayton:2008,nayak:2013,thesis:alok}; this leads
to unphysical wave dynamics. We check that our results are free
from such numerical artifacts by investigating the spatiotemporal
evolution of an expanding wave front that emerges from a point
stimulus. We find that fronts of the expanding wave do not
deviate substantially from circles, when we apply a point
stimulus of strength $450$~pA for $3$~ms at the center of the
domain.

We use two parameter sets P1 and P2 for myocytes to obtain,
respectively, a stable rotating spiral (RS) and a spiral-turbulence 
(ST) states in our 2D simulation domain, and a stable
rotating scroll or scroll-wave turbulence in our 3D domain. The
parameter set P1 is the original one used in the TP06
model~\cite{tp06,nayak:2013,thesis:alok}. In the P2 parameter
set, we use the following parameters, with all other parameters
the same as in the original TP06 model: (a) $G_{Kr}$, the
$I_{Kr}$ conductance, is $0.172$~nS/pF; (b) $G_{Ks}$, the
$I_{Ks}$ conductance, is $0.441$~nS/pF; (c) $G_{pCa}$, the
$I_{pCa}$ conductance, is $0.8666$~nS/pF; (d) $G_{pK}$, the
$I_{pK}$ conductance, is $0.00219$~nS/pF; and (e) $\tau_f$, the
time constant of the $f$ gating variable that is associated with the
$I_{CaL}$ current, is increased $2$ times compared to its value
in the TP06 model~\cite{tp06,nayak:2013,thesis:alok}. Our
fibroblasts parameters are as follows: $C_f=6.3$~pF, $G_f=4$~nS,
$E_f=-49.0$~mV, and $G_j$ in the range $0 \le G_j \le
6$~nS~\cite{nayak:2013,xie:2009a}.

To obtain spiral and scroll waves we use the S1-S2 cross-field
protocol~\cite{nayak:2014,majumder:book,thesis:alok}. We apply a
stimulus (S1) of strength $150$~pA for $3$~ms to the left
boundaries of our simulation domains, to form a plane wave. We
then apply the second (S2) stimulus, with the same strength and
duration as the S1 stimulus, from the bottom boundary and with
$0~{\rm mm} \leq y \leq 125~{\rm mm}$ in 2D, and $0~{\rm mm} \leq
y \leq 125~{\rm mm}$ and $0~{\rm mm} \leq z \leq 2~{\rm mm}$ in
3D. This protocol leads to the formation of spiral and scroll
waves, respectively, in our 2D and 3D domains.

To examine the spatiotemporal evolution of our system, we obtain
pseudocolor or isosurface plots of $V_m$, time series of $V_m$,
from representative points ($x = 125$ mm, $y = 125$ mm for 2D, and 
$x= 125$ mm, $y = 125$ mm, $z=1.25$ mm for 3D), which we mark with  an
asterisk ($\ast$) in all pseudocolor plots of $V_m$. We examine the
inter-beat interval (IBI), by using this time series with $4.4
\times 10^5$ data points; the IBI is the interval between two
successive spikes in this time series. We obtain the power
spectra $E(\omega)$, of the time series of $V_m$, by using $2
\times 10^5$ data points; to eliminate transients we remove the
initial $2.4 \times 10^5$ data points. 
To obtain the rotation period T of a spiral, in an RS state, we
average over the last $5$ rotations of that RS.

\section{Results}
\label{result}

In subsection~\ref{spiralwave}, we begin by studying spiral-wave
dynamics in a 2D domain of myocytes without fibroblasts; we then
introduce fibroblasts, either regularly or randomly, and examine
the effects they have on spiral-wave dynamics.
Subsection~\ref{scrollwave} contains the results of our studies
of scroll-wave dynamics in our 3D simulation domain.

\subsection{Spiral-wave dynamics in our 2D model}
\label{spiralwave}

\begin{figure}[!t]
\begin{center}
\includegraphics[width=\columnwidth]{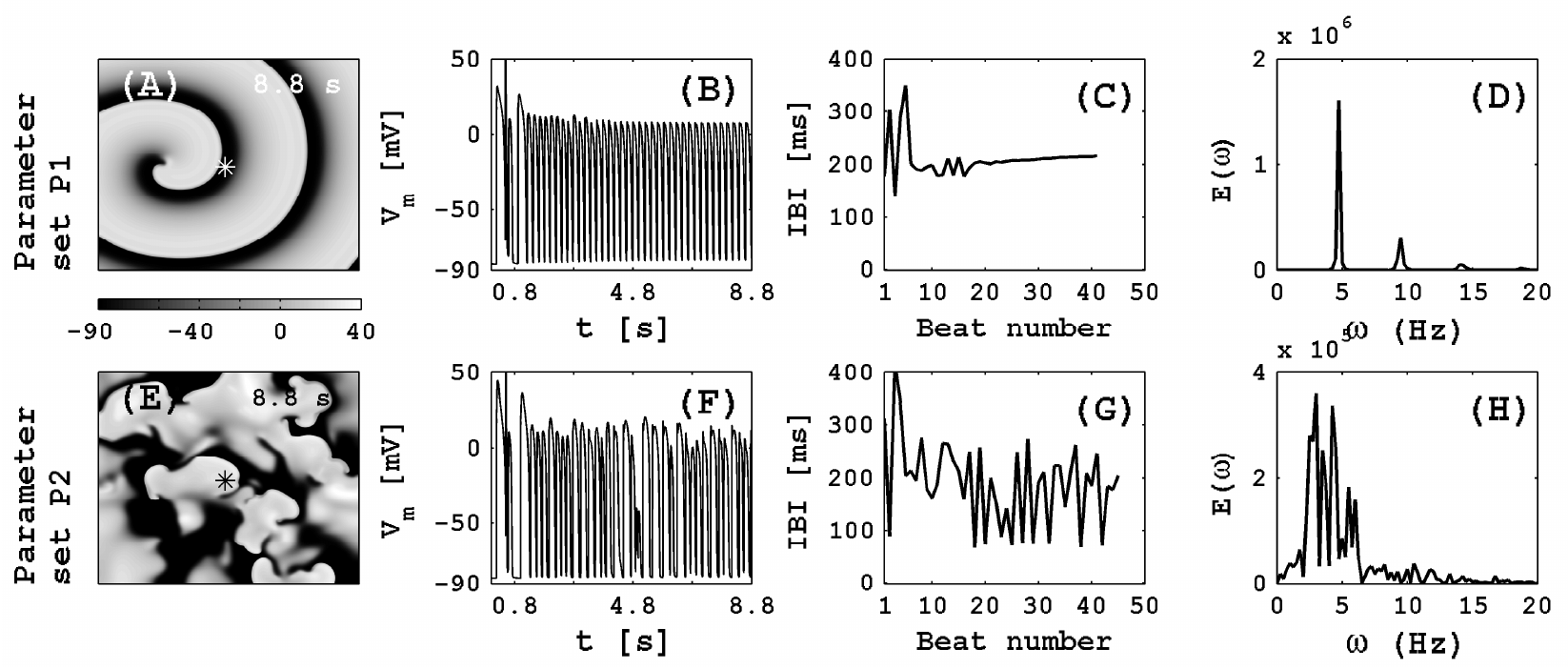}  
\caption{
The rotating-spiral (RS) and spiral-turbulence (ST) states in a
2D domain in the absence of fibroblasts. For the parameter set
P1, the pseudocolor plot of $V_m$ in (A), the periodic nature of
the local time series for $V_m$ from the representative point
(marked by an asterisk $\ast$ in (A)) in (B), the flattening IBI
with an average rotation period $T \simeq 212$~ms in (C), and the
discrete peaks in the power spectrum with the fundamental
frequency $\omega_f=4.75$~Hz and its harmonics in (D)
characterize the RS state. The exact analogs of plots (A)-(D) are
shown, respectively, in (E)-(H) for the P2 parameter set; the
irregular local time series, the fluctuating behavior of the IBI
and the broad-band nature of the power spectrum characterize the
ST state.}
\label{fig:myo2d}
\end{center}
\end{figure}

In Fig.~\ref{fig:myo2d}(A), we show a pseudocolor plot of $V_m$,
at time $t=8.8$~s, for the parameter set P1, in our 2D simulation
domain without fibroblasts; the initial condition evolves to a
state with a single rotating spiral (RS). The local time series of
$V_m(x,y,t)$, from the representative point shown by the asterisk
in Fig.~\ref{fig:myo2d}(A)), is given in Fig.~\ref{fig:myo2d}(B)
for $0~{\rm s}\le t \le 8.8~{\rm s}$; a plot of the IBI versus
the beat number is given in Fig.~\ref{fig:myo2d}(C), which shows
that, after initial transients, the spiral wave rotates
periodically with an average rotation period $T \simeq 212$~ms.
The power spectrum $E(\omega)$ in Fig.~\ref{fig:myo2d}(D) has
discrete peaks at the fundamental frequency $\omega_f\simeq
4.75$~Hz and its harmonics.  The periodic time series of $V_m$,
the flattening of the IBI, and the discrete peaks in $E(\omega)$
demonstrate that the time evolution of the spiral wave, with the
P1 parameter set, is periodic. In Figs.~\ref{fig:myo2d}(E)-(H),
we show the exact analogs of Figs.~\ref{fig:myo2d}(A)-(B) for
the P2 parameter set.  The non-periodic local time series in
Fig.~\ref{fig:myo2d}(F), the fluctuating IBI in
Fig.~\ref{fig:myo2d}(G), and the broad-band nature of $E(\omega)$
in Fig.~\ref{fig:myo2d}(H) are characteristic of the ST state.
The pseudocolor plot in Fig.~\ref{fig:myo2d}(E), at time
$t=8.8$~s, shows such an ST state, which arises from the steep slope of
the action-potential-duration-restitution (APDR)
plot~\cite{tp06,nayak:pre}. In summary, then, in the absence of
fibroblasts, the parameter sets P1 and P2 lead, respectively, to
(a) an RS state and (b) an ST state in our 2D simulation domain.

\begin{figure}
\begin{center}
\includegraphics[width=\columnwidth]{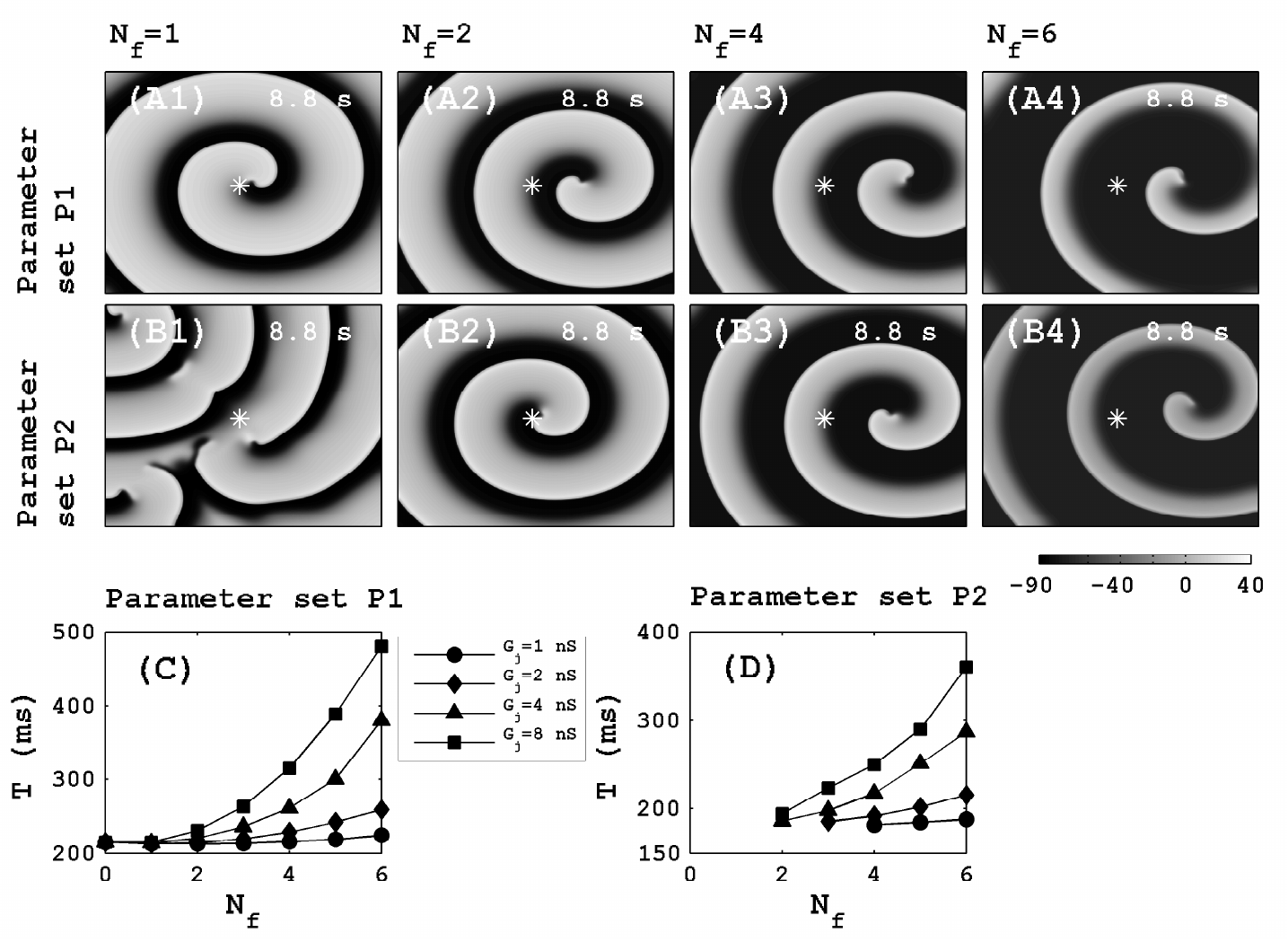}
\caption{
Various RS and ST states in our 2D domain for the regularly
attached fibroblast model. ((A1)-(B4)) Pseudocolor plots of
$V_m$, with $G_j=4$~nS for $N_f=1,\,2,\,4,{\rm and}~6$,
illustrate that the RS state (parameter set P1) remains in an RS
state as $N_f$ increases (first row).  However, an ST state
(parameter set P2) shows a transition to an RS state as $N_f$
increases (second row).  ((C)-(D)) The rotation period T of RS
increases as $N_f$ increases, for a fixed value of $G_j$, and
vice-versa, for both the P1 and P2 parameter sets.}
\label{fig:reg2d}	
\end{center}
\end{figure}

We now examine the effects of fibroblasts on spiral-wave dynamics
in both RS and ST states. We begin our investigation with the
P1 parameter set and with the regularly attached fibroblast model
with $1 \leq N_f \leq 6$ and $1~{\rm nS} \leq G_j \leq 8~{\rm
nS}$; the remaining parameters for fibroblasts are as in
subsection~\ref{method}.

In Figs.~\ref{fig:reg2d}(A1)-(A4) we show pseudocolor plots of
$V_m$, at time $t=8.8~$s, with the P1 parameter set for regularly
attached fibroblast model with $G_j=4$~nS and different values of
$N_f$. The analogs of Figs.~\ref{fig:reg2d}(A1)-(A4) are shown in
Figs.~\ref{fig:reg2d}(B1)-(B4) for the P2 parameter set. The
plots in the first row of Fig.~\ref{fig:reg2d} show that the RS
state, which we obtain in the absence of fibroblasts, does not
evolve into an ST state; however, the spiral-arm width $W_d$
decreases as we increase $N_f$, for a fixed value of $G_j$ (first
row of Fig.~\ref{fig:reg2d}).  We define $W_d$ to be the
difference of the radial distance between the wave front and the
wave back of a spiral arm, whose center is located at the spiral
core. Such a decrease of $W_d$ is related to the shortening of
the action-potential-duration (APD) of a myocyte-fibroblast
composite that has been discussed in
Refs.~\cite{nayak:pre,nayak:2013}. 

For the P2 parameter set, we observe a transition from an ST to
an RS state as we increase $N_f$ for a fixed value of $G_j$
(second row of Fig.~\ref{fig:reg2d}). Such an ST-RS transition is
the consequence of the suppression of the steep APDR slope of a
myocyte-fibroblast composite at the cellular
level~\cite{nayak:pre,petrov:2010}. Once the ST state is
suppressed, a single spiral in an RS state rotates periodically
as shown Figs.~\ref{fig:reg2d}((B2)-(B4)). In Figs.~\ref{fig:reg2d}(C) and
(D), we plot, respectively, the rotation period T of a spiral
wave in an RS state versus $N_f$, for different values of $G_j$,
for the P1 and P2 parameter sets.  We find that T increases as (i) $N_f$
increases, with a fixed value of $G_j$, and (ii) $G_j$ increases,
with a fixed value of $N_f$.  This increase of T is a
consequence of the decrease of CV that is associated with a
decrease of the upstroke velocity of a myocyte-fibroblast
composite AP in its depolarization phase, as shown in
Refs.~\cite{nayak:2013,nayak:pre,xie:2009}.  Furthermore, we
observe that the minimum value of $N_f$, required for the ST-RS
transition, decreases as $G_j$ increases, for the P2 parameter
set.

\begin{figure}[!t]
\begin{center}
\includegraphics[width=\columnwidth]{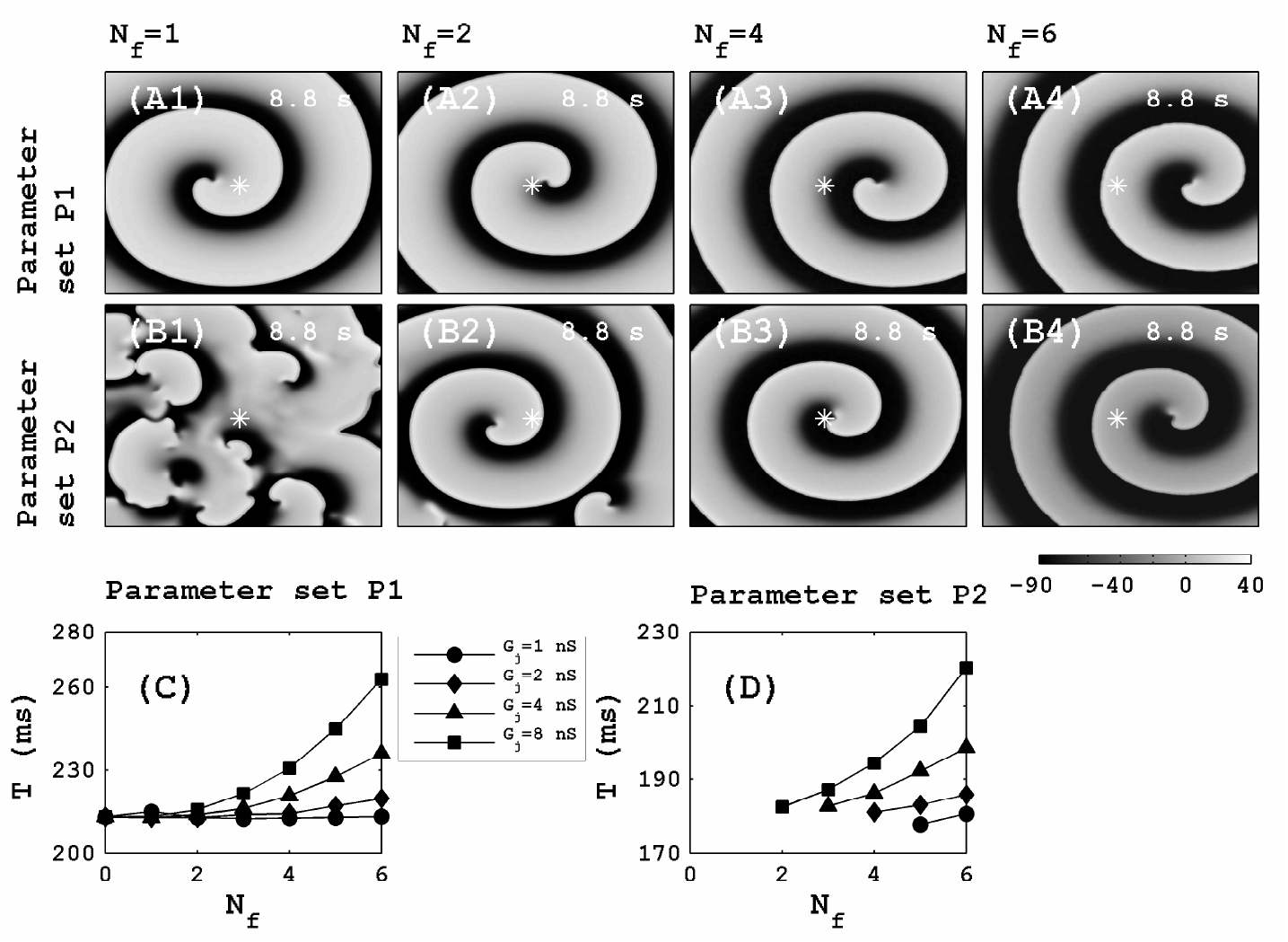}  
\caption{
Various RS and ST states in our 2D domain for the randomly
attached fibroblast model (this figure is the analog of
Fig.~\ref{fig:reg2d}). The results are qualitatively similar to
those in Fig.~\ref{fig:reg2d} for the regularly attached
fibroblast case.  Note that the minimum value of $N_f$, for a fixed
value of $G_j$, for the ST-RS transition, is higher compared to
that in Fig.~\ref{fig:reg2d}(D).}
\label{fig:rand2d}
\end{center}
\end{figure}

We focus next on spiral-wave dynamics, with P1 and P2 parameter
sets, in our randomly attached fibroblast model.  In
Fig.~\ref{fig:rand2d}, we show the exact analogs of
Fig.~\ref{fig:reg2d}, but now for the randomly attached
fibroblast model.  The pseudocolor plots in
Figs.~\ref{fig:rand2d}(A1)-(A4) show that the randomness in
attaching fibroblasts does not lead to an RS-ST transition for
the P1 parameter set. However, inspite of the randomness in the
arrangement of fibroblasts, we observe an ST-RS transition for
the P2 parameter set (see Figs.~\ref{fig:rand2d}(B1)-(B4)), which
is qualitatively similar to the ST-RS transition in the P1 case
(compare the second rows of Figs.~\ref{fig:reg2d} and
\ref{fig:rand2d}). However, the minimum value of $N_f$, required
for an ST-RS transition, is higher for the randomly attached
fibroblast model than in the regularly attached case (compare
Figs.~\ref{fig:reg2d}(D) and \ref{fig:rand2d}(D)). 

\begin{figure}
\begin{center}
\includegraphics[width=\columnwidth]{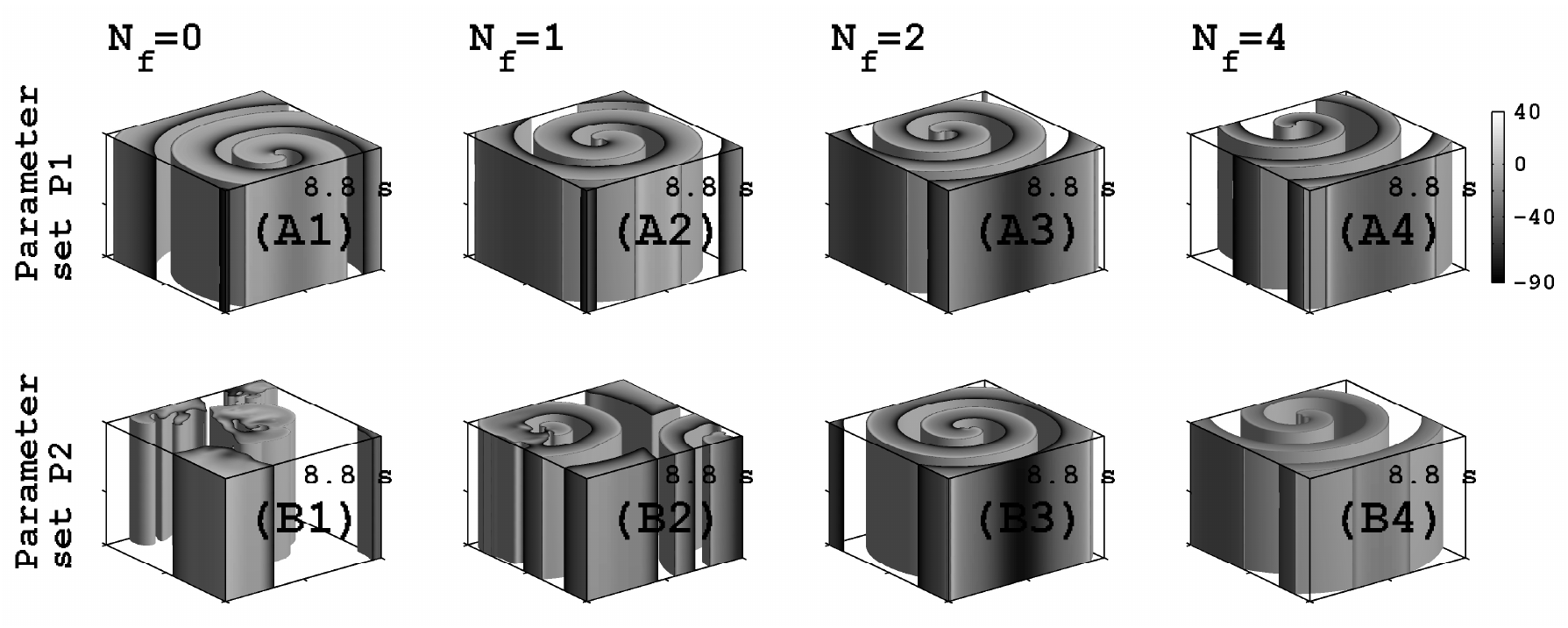}
\includegraphics[width=\columnwidth]{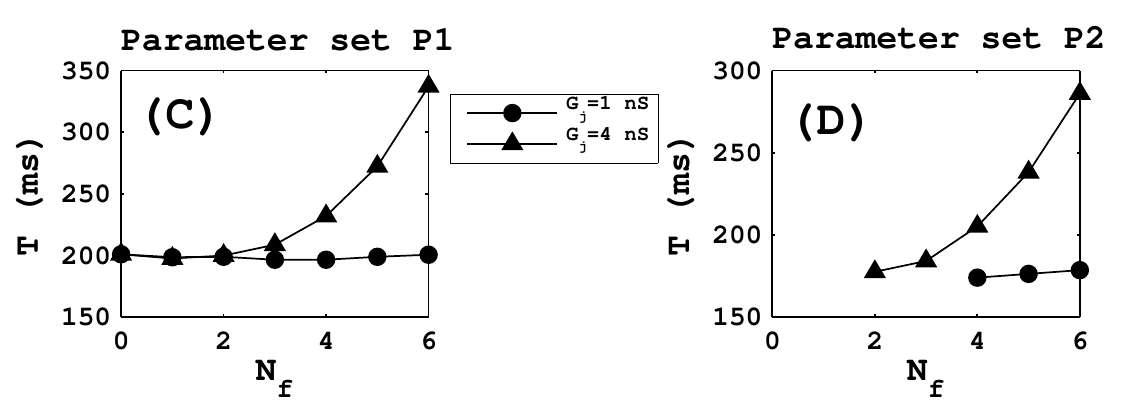}
\caption{
The rotating-scroll and scroll-wave-turbulence states, in our 3D
simulation domain of size $256\times256\times2$~mm$^3$, for the
regularly attached fibroblast model, are shown in ((A1)-(B4)) via
isosurface plots of $V_m$. The myocyte-fibroblast coupling
strength $G_j=4$~nS. The scroll-arm width of a rotating scroll,
with the P1 parameter set, decreases as $N_f$ increases (first
row). The scroll-wave turbulence, associated with the P2
parameter set, is converted to a rotating scroll as $N_f$
increases (second row). ((C)-(D)) Plots of the rotation period T
of a scroll wave in a rotating-scroll state, for the P1 and P2
parameter sets; for both parameter sets T increases as $G_j$
increases; note that, for the ST-RS transition, the minimum value
of $N_f$ decreases as $G_j$ increases.}
\label{fig:reg3d}
\end{center}
\end{figure}

\begin{figure}
\begin{center}
\includegraphics[width=\columnwidth]{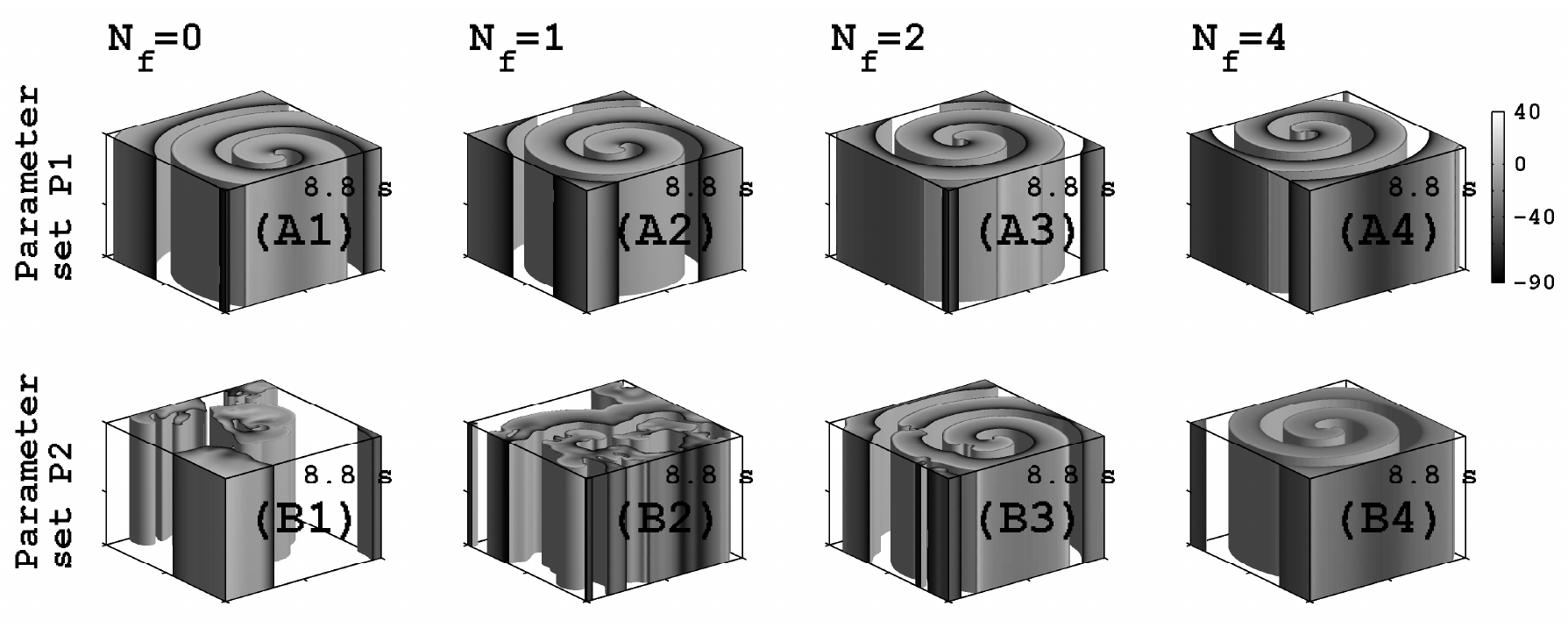}
\includegraphics[width=\columnwidth]{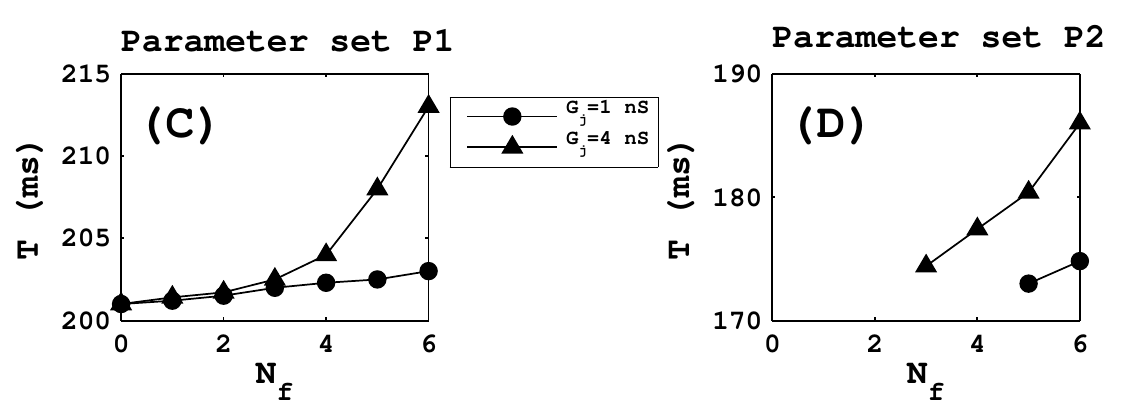}
\caption{
The rotating-scroll and scroll-wave-turbulence states, in our 3D
simulation domain for the randomly attached fibroblast model; the
exact analog of Fig.~\ref{fig:reg3d}. The results are
qualitatively similar to those for the regularly attached
fibroblast case.  Note that the ST-RS transition $N_f$ value, for a
fixed value of $G_j$, is higher than its counterpart in
Fig.~\ref{fig:reg3d}(D).}
\label{fig:rand3d}
\end{center}
\end{figure}

\subsection{Scroll-wave dynamics in our 3D model}
\label{scrollwave}

We turn now to a systematic study of scroll-wave dynamics in our
3D simulation domain. For both the P1 and P2 parameter sets and
both regularly and randomly attached fibroblast models, we carry
out simulations to study the dependence of scroll-wave dynamics
on $N_f$ and $G_j$. We present our numerical results below.

In Figs.~\ref{fig:reg3d}(A1), (A2), (A3) and (A4), we show,
respectively, isosurface plots of $V_m$, at time $t=8.8$~ms, for
the P1 parameter set in our regularly attached fibroblast model
with $G_j=4$~nS and $N_f=0$ (i.e., isolated myocytes), $N_f=1$,
$N_f=2$, and $N_f=4$.  In the absence of fibroblasts, i.e.,
$N_f=0$, the P1 parameter set displays a rotating scroll wave
with fundamental frequency $\omega_f \simeq 5$~Hz and rotation
period T $\simeq$ 201~ms; this is consistent, because
$\omega_f\simeq1/{\rm T}$. In Fig.~\ref{fig:reg3d}(C), we plot T
versus $N_f$ for $G_j=$~1 nS ($\bullet$) and 4~nS
($\filledtriangleup$). We find that T increases as we increase
(i) $N_f$, for a fixed value of $G_j$, or (ii) $G_j$, for a fixed
value of $N_f$. In Figs.~\ref{fig:reg3d}((B1)-(B4)) and (D), we
show, respectively, the exact analogs of
Figs.~\ref{fig:reg3d}((A1)-(A4)) and (C), for the P2 parameter
set. In the absence of fibroblasts and for the P2 parameter set,
we obtain a scroll-wave-turbulence state
(Fig.~\ref{fig:reg3d}(B1)); this scroll-wave turbulence is
converted to a rotating scroll if we have $N_f >1$ (second row of
Fig.~\ref{fig:reg3d}). Once the scroll-wave turbulence state is
suppressed, a rotating scoll rotates with a period T, which
increases as we increase $N_f$ for a fixed value of $G_j$, and
vice-versa (Fig.~\ref{fig:reg3d}(D)).  Furthermore, from
Fig.~\ref{fig:reg3d}(D), we find that the minimum value of $N_f$,
required for the  ST-RS transition, is 4 and 2, respectively, for
$G_j=$ 1~nS ($\bullet$) and 4~nS ($\filledtriangleup$). The
isosurface plots in Fig.~\ref{fig:reg3d} show that the width
$W_d$ of a scroll-wave arm in the rotating-scroll state decreases
as we increase $N_f$ for both the P1 and P2 parameter sets. The
mechanisms of the ST-RS transition, and increase of T and a decrease of
$W_d$, as we increase $N_f$ and $G_j$, are the same as those
we have found in our 2D studies.

In Fig.~\ref{fig:rand3d} we show the exact analog of
Fig.~\ref{fig:reg3d} for the randomly attached fibroblast model,
with both the P1 and P2 parameter sets. Our
scroll-wave results here are similar to those for the case of
regularly attached fibroblast model.  From
Fig.~\ref{fig:rand3d}(D), we find that the minimum value of $N_f$,
for the ST-RS transition, is 5 and 3, respectively, for $G_j=1$~nS
($\bullet$) and 4~nS ($\filledtriangleup$). Note that this
minimum value of $N_f$ is higher for the randomly attached
fibroblast model than it is for the regularly attached
fibroblast model (compare Figs.\ref{fig:reg3d}(D) and
\ref{fig:rand3d}(D)).

\section{Conclusions}
\label{conclusion}

We have presented the most extensive numerical study carried out
so far of the effects of fibroblasts on spiral- and scroll-wave
dynamics in a mathematical model for human ventricular tissue with
fibroblasts, attached regularly or randomly to myocytes.  Our
numerical study has been designed to uncover the role of (i) the
organization of fibroblasts in ventricular tissue ( i.e., to
compare regular and random arrangements), (ii) myocyte-fibroblast
coupling $G_j$, and (iii) the density of fibroblasts, i.e.,
the maximum number of fibroblasts $N_f$ attached to a myocyte.  One
of the principal results of our studies is that spiral- and
scroll-wave dynamics depend only slightly on the details of the
organization of fibroblasts in ventricular tissue. However, the
ST-RS transition, the stability of spiral- and scroll-wave
turbulence, the rotation period of a rotating spiral and scroll,
and the width of a rotating spiral and scroll arms, depend
sensitively on $N_f$ and $G_j$. 

Earlier studies have investigated the effects of fibroblasts on
spiral-wave dynamics by introducing randomly diffuse fibroblasts
in a myocyte domain~\cite{zlochiver:2008,majumder:2012}. Such
randomly diffuse fibroblasts in a myocyte domain inhibit
electrical-wave propagation, and initiate spiral-wave turbulence
state. Studies by Xie, \textit{et al}~\cite{xie:2009} have found
that spiral-wave breakup occurs, in an LR1 model, because of randomly
diffuse fibroblasts in a localized area of a simulation domain;
Zlochiver, \textit{et al.}~\cite{zlochiver:2008} have shown from
their experiments and simulations that a rotating spiral becomes
unstable and, finally, spiral breakup occurs, as they increase the
percentage of diffuse fibroblasts.  Majumder, \textit{et
al.}~\cite{majumder:2012} have shown from their numerical
experiments that a transition from an RS to various ST states
occurs depending on the percentage of fibroblasts in their
simulation domain. In our attached-fibroblast model studies,
fibroblasts do not inhibit wave propagation~\cite{nayak:2013};
however, fibroblasts attached to a myocyte can lower the
steepness of the APDR curve, depending on the values of $G_j$ and
$N_f$~\cite{nayak:pre,petrov:2010}. Such a lowering of the steep
slope of the APDR eliminates spiral- and scroll-wave turbulence
in our 2D and 3D simulation
domains~\cite{qu:2000,qu:2000a,weiss:2000,garfinkel:2000}.
Therefore, we observe an ST-RS transition.  Earlier studies in
Ref.~\cite{petrov:2010} have observed ST-RS spiral-wave
transitions because of a suppression of the steep portion of the
APDR slope in a 3D model consisting of myocytes, fibroblasts, and
extracellular space by using the LR1 model~\cite{lr1}. However,
those studies have not investigated the spiral- and scroll-wave
transition as a function of $N_f$ and $G_j$. Our study shows that
both $N_f$ and $G_j$ are important factors during the fibrosis
process~\cite{nguyen:2014,biernacka:2011}. 

We suggest that our results from \textit{in silico} studies can
be verified in \textit{in vitro} experiments. Furthermore, by
using advanced cell-culture
techniques~\cite{haraguchi:2012,shimizu:2002,baudino:2008}, our
2D and 3D numerical results can be tested easily in
\textit{cell-culture} experiments.

\section*{Acknowledgments}
We thank the Department of Science and Technology (DST), India,
the University Grants Commission (UGC), India, and the Robert
Bosch Centre for Cyber Physical Systems (RBCCPS), IISc, for
support.

\bibliographystyle{apsrev} 
\bibliography{references}

\end{document}